\documentclass[prb,twocolumn,showpacs,preprintnumbers,amsmath,amssymb]{revtex4}
\usepackage{graphicx}
\usepackage{dcolumn}
\usepackage{bm}

\begin{document}

\title{Continuum theory for the piezoelectric response of chiral nanotubes
    under uniaxial and torsional stresses}
\author{P. J. Michalski}
\author{E. J. Mele}

\affiliation{Department of Physics and Astronomy, University of
Pennsylvania, Philadelphia, PA 19104}

\date{\today}

\begin{abstract}
We develop and solve a continuum theory for the piezoelectric
response of nanotubes under applied uniaxial and torsional stresses.
We find that the piezoelectric response is controlled by the chiral
angle, the aspect ratio, and two dimensionless parameters specifying
the ratio of the strengths of the electrostatic and elastic
energies.  The model is solved in two limiting cases and the
solutions are discussed.  These systems are found to have several
unexpected physical effects not seen in conventional bulk systems,
including a strong stretch-twist coupling and the development of a
significant bound charge density in addition to a surface charge
density.  The model is applied to estimate the piezoelectric
response of a boron-nitride nanotube under uniform tensile stress.
\end{abstract}

\pacs{77.65.Ly, 73.63.Bd, 73.63.Fg}


\maketitle

\section{\label{sec1} Introduction}

Nanotubes and nanowires are a family of one dimensional structures
whose properties can be tailored by their chemical composition,
size and shape. The lateral dimensions of these structures can be
reduced to the nanometer scale, a feature that is exploited in
almost all applications envisioned for these objects. Carbon
nanotubes are the prototypical member of this family; they are the
simplest chemically and are known to exhibit a diversity of
electronic behaviors that are determined by the boundary
conditions imposed on their electronic wavefunctions when they are
mapped onto a cylindrical surface. While this sensitivity makes
for rich theoretical study, it also makes CNTs difficult to use in
applications since tubes must be identified and sorted according
to their lattice structures. Boron-nitride nanotubes (BNNTs) are
the III-V analog of CNTs and are an attractive alternative to CNTs
because they have similar mechanical properties, are chemically
inert over a wide range of conditions, and importantly, are all
semiconducting with a band gap that is nearly independent of the
chirality.

A BN nanotube is a heteropolar structure with a lower lattice
symmetry than the carbon nanotube. Importantly,  a twofold rotation
about a perpendicular axis through the tube interchanges its A and B
sublattices; this is a symmetry operation of the carbon nanotube but
not of the boron nitride tube. In early theoretical work it was
appreciated that this allows for a nonzero electric polarization in
the ground state of the heteropolar tube \cite{mk02} and a linear
coupling of the polarization to a uniform strain \cite{ncmbn03},
i.e. BN nanotubes are piezoelectric \cite{sm03}. Interestingly, the
piezoelectricity is a property inherited from the flat BN sheet,
while the electric polarization of an unstrained tube is a new
property that only occurs when the sheet is wrapped to form a
nanotube.

The description of piezoelectricity in a one dimensional nanotube
differs fundamentally from the analogous formulation for a three
dimensional crystal. In an ordinary three dimensional piezoelectric
material described within the standard Landau-Devonshire theory, its
macroscopic polarization (a vector) is linearly coupled to a uniform
strain field (a tensor) through a third-rank piezoelectric tensor.
Physically, this reflects the fact that bound charges arise from a
divergence of the polarization, and these bound charges are mutually
electrostatically coupled by the Coulomb interaction which diverges
at small momentum proportional to $1/q^2$. Thus the free energy
contains a term that is bilinear in the polarization that remains
constant in the long wavelength $q \rightarrow 0$ limit. As a
consequence, a uniformly strained three dimensional piezoelectric
induces a state of uniform bulk polarization and the bound charges
reside precisely at the sample boundaries.

This state of affairs is upset for a one dimensional piezoelectric
because the Coulomb potential that couples the bound charges has
only a softer $-\log q$ long wavelength divergence in one dimension.
Thus the analogous electrostatic coupling bilinear in the
polarization {\it vanishes} in the long wavelength limit. The
consequences of this are quite interesting and were partially
explored in our previous work \cite{pnm05}. In essence the strain
and polarization must be treated as inhomogeneous fields, whose
equilibrium forms can be obtained by minimization of an appropriate
free energy that contains a nonlocal coupling between the one
dimensional strain gradients. As a consequence, a uniform applied
mechanical stress generally induces a state of nonuniform strain in
a nanotube or nanowire, and the bound charge redistributes from the
tube end into its interior in a self consistent fashion. The
distribution of the equilibrium strain and polarization fields
depends on a subtle interplay of their elastic and electrostatic
interactions. Experimentally available nanotubes are expected to
fall in a parameter regime where the elastic interactions dominate
the problem so that most real materials, including BN nanotubes, are
elastically ``stiff."

A second rather interesting consequence of this nonlocal theory
occurs when different strain fields are separately coupled to the
electric polarization. Elimination of the polarization then induces
a bilinear coupling between different strain gradients. A striking
consequence of the induced coupling occurs on a nanotube where both
tube extension (longitudinal strain) and tube torsion (transverse
strain) couple to the polarization field. Here, a uniform tensile or
compressive stress induces extension or compression {\it and} a
concomitant torsion in its equilibrium state. This conversion of a
longitudinal to a torsion stress is impossible in a macroscopic
homogeneous tube or cylinder, though it would generically occur in
the elastic theory of a helical filament. We conclude that the
appearance of an anomalous elastic response of this type is a
macroscopic manifestation of the underlying (atomic scale) chirality
of the structure.

In this paper we develop a theory for this anomalous elastic
response and apply it to study the equilibrium strained state of a
family of BN nanotubes. We find that it has a rather rich structure,
depending on the relative strengths of the elastic and electrostatic
interactions and on the chiral angle of the tube. Nonetheless, these
anomalous couplings turn out to be fairly small, and will likely be
difficult to observe directly in ordinary experimental environments.
Physically, this reflects the fact that finite radius BN are
mechanically stiff structures that tend to be dominated by their
bare elastic interactions, weakly perturbed by the electrostatic
effects of the type considered here.

The rest of this paper is organized as follows.  In
Sec.~\ref{sec2} we develop a continuum theory appropriate for
$1$-D systems and discuss limiting cases.  In Sec.~\ref{sec3} we
present exact numerical and approximate analytic solutions for the
various parameter regimes.  In Sec.~\ref{sec4} we apply our
results to chiral BNNTs and calculate their piezoelectric
response. Finally, in Sec.~\ref{sec5} we summarize our results and
present conclusions.

\section{\label{sec2} Free Energy Formalism}

We consider a piezoelectric nanotube with radius $R$ and length
$L$, and let the tube axis be the $z$-axis.  In what follows, we
assume that $L \gg R \gg a_0$, where $a_0$ is the B-N lattice
spacing.  In this limit of large aspect ratio and large tube
radius a continuum theory is appropriate.

The tube is placed under a constant external stress, which we
partition into its longitudinal component $f_s$  and torsional
component $f_t$. The applied stresses induce  strain fields
$\eta_s (z)$ (stretch) and $\eta_t (z)$ (twist). The induced
polarization is related to these strains through $P(z) = e_s
\eta_s (z) + e_t \eta_t (z)$, where $e_s$ and $e_t$ are the linear
piezoelectric constants that depend on the chiral angle $\theta$.
For a general chiral nanotube both $e_s$ and $e_t$ are nonzero.
For the two high symmetry classes of nanotubes only one
piezoelectric coefficient is nonzero, $e_s$ for ``zigzag''
nanotubes and $e_t$ for ``armchair'' nanotubes.  The special cases
where only one elastic degree of freedom couples to the
polarization have no induced cross coupling terms and were
considered previously.\cite{pnm05}

The free energy of the system can be written as the sum of three
terms,
\begin{equation}\label{eq00}
G = G_{\rm elastic} + G_{\rm electro.} - \sum_i \int_0^L f_i
\eta_i (z) dz ,
\end{equation}
where $G_{\rm elastic}$ is the elastic contribution, $G_{\rm
electro.}$ is the electrostatic contribution, and the last term
represents the work done by the applied stress, and the sum runs
over $s$ and $t$. The elastic contribution is
\begin{equation}\label{eq01}
G_{\rm elastic} = \sum_{i,j} \frac{1}{2} C_{ij} \int_0^L \eta_i(z)
\eta_j(z) dz \hspace{1 mm},
\end{equation}
where the indices $i$ and $j$ run over $s$ and $t$ and we have
introduced four $1$-D elastic moduli $C_{ij}$. The diagonal $1$-D
elastic moduli are radius dependent quantities with units of
energy/length, and are given by $C_{ss} = 2\pi R C_{11}$ and $C_{tt}
= 2 \pi R C_{66}$. $C_{11}$ and $C_{66}$ are $2$-D elastic constants
in the conventional Voigt notation and are the relevant intensive
variables, with units energy/area, which depend on the nanotube
chemistry.  The off-diagonal term, $C_{st}$, is small and decreases
as $1/R^2$ in the limit of large radius tubes. We therefore ignore
this term by setting $C_{st} = 0$ in all that follows.

Using the fact that the bound charge density is related to spatial
derivatives of the strains through $\rho(z) = -\sum_i e_i
\partial \eta_i (z)/\partial z$, the electrostatic contribution to the free energy can be
written as
\begin{equation} \label{eq02} G_{\rm electro.} =
\sum_{i,j} \frac{e_i e_j}{2}  \int_0^L \!\!\! \int_0^L \!\!
\frac{\partial \eta_i(z)}{\partial z} V(z-z^\prime) \frac
{\partial \eta_j (z^\prime)}{\partial z^\prime} \,\, dz \,
dz^\prime \hspace{1 mm} ,
\end{equation}
where again the sums on $i$ and $j$ are over $s$ and $t$. Here,
$V(z-z^\prime)$ is a Coulomb kernel describing the electrostatic
interaction between rings of charge centered at positions $z$ and
$z^\prime$ along the tube axis.  $e_s$ and $e_t$ are $1$-D
piezoelectric constants with units of charge. They may be obtained
by rotating the piezoelectric constants for an infinite $2$-D sheet
onto the symmetry axes of a tube.\cite{sm03} The three-fold symmetry
of an infinite hexagonal lattice reduces the number of independent
$2$-D piezoelectric constants to one, which we take as $e_2 \equiv
e_{xxx}$ with units of charge/length.
 If we define the chiral angle as the angle between the nanotube
axis and a $2$-D primitive translation vector ($\theta = 30^\circ$
is the zigzag configuration and $\theta = 0^\circ$ is the armchair
configuration), then the $1$-D piezoelectric constants are given
by
\begin{eqnarray}\label{eq03}
e_s &=& C e_2 \sin(3\theta) \nonumber \\
e_t &=& - C e_2 \cos(3\theta) \hspace{1 mm} ,
\end{eqnarray}
where $C$ is the nanotube circumference.\cite{sm03}  These
expressions are derived neglecting finite radius corrections due
to tube curvature.  However, these expressions agree with
numerical results to within $15\%$ for the smallest nanotubes
considered, with better agreement for larger radius tubes where
curvature effects are reduced.

The polarization induced along the tube is composed of dipoles with
both a longitudinal and an azimuthal component.  The longitudinal
component accounts for the electrostatic energy given in
Eq.~(\ref{eq02}).  The total dipole moment of the azimuthal
polarization integrates to zero, but higher order moments do not
vanish and may contribute to the energy.  The energy from these
interacting rings of dipoles is of the form
\begin{equation}
G_{\rm rings} \sim -\left(\frac{a_0}{R}\right)^2 \frac{e_i^\phi
e_j^\phi}{R} \int \!\!\! \int dx dx^\prime \eta_i (x)
K(x-x^\prime) \eta_j (x^\prime) ,
\end{equation}
where the integral is written in terms of the dimensionless
variable $x = z/R$ and $e_i^\phi$ is the appropriate piezoelectric
constant giving the local azimuthal polarization in terms of the
induced strains.  The important result is that the kernel goes
asymptotically as $K(x) \sim 1/x^{2 m + 1}$, where $m$ is
approximately the number of $2$-D unit cells in one tube
circumference.  This kernel decays so fast that we may approximate
it with a delta function.  This term then has the same functional
form as $G_{\rm elastic}$. However, as we will see below, the
terms we will keep in $G_{\rm elastic}$ scale with the radius as
$R^2$, whereas all terms in $G_{\rm rings}$ scale as $1/R^3$.
Since we are interested in the limit of large radius tubes, this
energy represents a minor correction to the elastic energy of the
system and will be ignored.

The strain field may be discontinuous at the boundaries of the
nanotube.  To account for this, we let $\eta_i (z) = g_i (z) H(z)$,
where $g_i(z)$ is a continuous function of $z$ and $H(z) = 1$ for $0
\leq z \leq L$ and $0$ otherwise.  Writing $\eta_i(z)$ in this form
is convenient because it allows us to keep track of the delta
function contributions to $\eta_i^\prime(z)$ at the boundary of the
nanotube. In particular, it removes any ambiguity that might arise
in an expression such as $\eta_i^\prime(0)$. In what follows we will
call both $\eta_i(z)$ and $g_i(z)$ the strain field, since they
agree over the length of the tube.

We extract the dimensional dependence of the energies in
Eq.~(\ref{eq00}) by expressing all lengths in units of the tube
radius and all radius dependent quantities by their relevant
intensive quantities. We define a scaled length $x = z/R$ and the
dimensionless aspect ratio as $\lambda_1 = L/R$.  Then, after
integrating the electrostatic free energy by parts, the free energy
is given by
\begin{widetext}
\begin{eqnarray}\label{eq05}
G & = & \pi R^2 \int_0^{\lambda_1}\!\!\! dx [C_{11} g_s^2 (x) +
C_{66} g_t^2 (x)] - R \int_0^{\lambda_1}\!\!\! dx[f_s g_s (x) + f_t
g_t(x)] \nonumber
\\ & & + 2 \pi^2 R e_2^2 \int_0^{\lambda_1}\!\!\!dx dx^\prime \frac{\partial^2
V(x-x^\prime)}{\partial x \partial x^\prime} \left[\sin^2(3\theta)
g_s(x) g_s(x^\prime) - 2 \cos(3\theta) \sin(3\theta) g_s(x) g_t
(x^\prime) + \cos^2(3\theta) g_t(x) g_t (x^\prime)\right] \nonumber
\\ & & - 4 \pi^2 R e_2^2 V^{\prime} (0+) \int_0^{\lambda_1}\!\!\! dx [\sin^2(3\theta) g_s^2(x) - 2 \sin(3\theta) \cos(3\theta) g_s(x)
g_t(x) + \cos^2(3\theta) g_t^2(x)] \hspace{1 mm},
\end{eqnarray}
\end{widetext}
where $V^\prime(0+)$ indicates the one-sided right derivative of the
Coulomb kernel.  It may be that $V^{\prime}(0+)=\infty$, but there
is an equal and opposite infinity in the double integral so that the
energy always remains finite.

The equilibrium strain fields are obtained by minimizing $G$ with
respect to the strains and satisfy
\begin{equation}\label{eq06}
\vec{g}(x) - 2V^{\prime} (0+) \Lambda \vec{g}(x) +
\int_0^{\lambda_1} \!\!\!dx^\prime \frac{\partial^2
V(x-x^\prime)}{\partial x
\partial x^\prime} \Lambda \vec{g}(x^\prime) = \vec{\beta} .
\end{equation}
For notational simplicity we have defined a strain ``vector'' as
$\vec{g}(x) = (g_s(x), g_t(x))$, and an elastic limit ``vector'' as
$\vec{\beta} = (\beta_s, \beta_t)$, where $\beta_s = f_s/(2\pi R
C_{11})$ is the value of the uniaxial strain in the absence of
piezoelectric effects and $\beta_t = f_t/(2 \pi R C_{66})$ is the
value of the torsional strain in the absence of piezoelectric
effects.  The matrix $\Lambda$ is given by
\begin{equation}\label{eq07}
\Lambda = 2\pi \left(\begin{array}{cc} \lambda_2 \sin^2(3\theta) &
-\lambda_2 \sin(3\theta) \cos(3\theta) \\ -\lambda_3 \sin(3\theta)
\cos(3\theta) & \lambda_3 \cos^2 (3\theta) \end{array} \right) ,
\end{equation}
where we have introduced two dimensionless constants, $\lambda_2 =
e_2^2 / (R C_{11})$ and $\lambda_3 = e_2^2 / (R C_{66})$, which
measure the relative strengths of the electrostatic and elastic
response.  Note that $C_{11}$ and $C_{66}$ are generally of the
same order of magnitude, so $\lambda_2$ and $\lambda_3$ will also
be of the same order of magnitude.

We will see below that the solutions to Eq.~(\ref{eq06})
asymptotically approach the elastic limit with a characteristic
decay length controlled by $\lambda_2$ and $\lambda_3$.  There are
two limiting cases of interest.  When the tube is much longer than
the decay length ($\lambda_1 \gg 2\pi\lambda_2$,$2\pi\lambda_3$),
the strains are near their elastic limits over most of the tube
length, with deviations from the elastic limit near the tube ends.
Such systems are dominated by their elastic response to the applied
stress, and we call this the elastically dominated limit.
Conversely, when the tube is much shorter than the decay length
($\lambda_1 \ll 2\pi\lambda_2$,$2\pi\lambda_3$), the strains never
decay to their elastic limits.  Such systems are dominated by their
electrostatic response to the applied stress, and we call this the
electrostatically dominated limit.  Solutions for these two limiting
cases will be explored in the next section.

The matrix $\Lambda$ is not invertible, therefore the strains are
not linearly independent.  We find that the strains are related by
\begin{equation}\label{eq08}
g_t (x) = \beta_t + \frac{\lambda_3 \cos(3\theta)}{\lambda_2
\sin(3\theta)} \left[\beta_s - g_s(x)\right] \hspace{1 mm} .
\end{equation}
Plugging this relationship into equation (\ref{eq06}) allows us to
write the equilibrium equation in terms of $g_s (x)$ alone,
\begin{eqnarray}\label{eq09}
g_s (x) &  = & \beta_s - 2V^{\prime} (0+) \left[ \gamma g_s (x) -
\tau \right] \nonumber\\&& + \int_0^{\lambda_1}\!\!\!dx^\prime
\frac{\partial^2 V(x-x^\prime)}{\partial x
\partial x^\prime} \left[\gamma g_s (x^\prime) - \tau\right]
\hspace{1 mm} ,
\end{eqnarray}
where we have defined two new constants,
\begin{equation}\label{extraeq1}
\gamma = 2 \pi (\lambda_2 \sin^2(3\theta) + \lambda_3
\cos^2(3\theta))
\end{equation}
and
\begin{equation}\label{extraeq2}
\tau = 2 \pi \cos(3\theta) [\lambda_3 \cos(3\theta) \beta_s +
\lambda_2 \sin(3\theta) \beta_t].
\end{equation}

In order to solve Eq.~(\ref{eq09}) an appropriate Coulomb kernel
must be specified.  In the simplest continuum model where the
kernel describes the  interaction of charged rings at positions
$x$ and $x'$ on a tube, the Coulomb kernel is given by
\begin{equation}\label{eq10}
V(x-x^\prime)=\frac{1}{2\pi} \int_0^{2\pi}\!\!\!\!
\frac{d\theta}{\sqrt{(x-x^\prime)^2 + \sin^2(\theta) +
(1-\cos(\theta))^2}} \hspace{1 mm} .
\end{equation}
This kernel diverges logarithmically as $x \rightarrow x^\prime$
and leads to an infinite energy for surface charges.  Using this
kernel the boundary condition on the strain fields is $e_s
\eta_s(0) + e_t \eta_t (0) = 0$.  A divergent energy is an
artifact of our continuum model and signals the breakdown of the
model at short distances.  It is useful to replace the divergent
kernel with a softened kernel,
\begin{equation} \label{eq11}
V_{\rm soft}(x-x^\prime) = \frac{1}{|x-x^\prime| + \alpha}
\hspace{1 mm} ,
\end{equation}
which retains the correct long range behavior but remains finite
as $x \rightarrow x^\prime$.  This kernel allows the presence of a
nonzero surface charge, with the ``cost'' of a surface charge
controlled by the cutoff $\alpha$.  The tube radius forms a
natural length scale below which the continuum model breaks down,
suggesting $\alpha \approx 1$ is an appropriate cutoff.  In
Ref.~(\onlinecite{pnm05}) it was shown that using the softened
kernel with $\alpha = 1$ gave nearly identical results for the
strain, charge density, and potential across the tube as those
obtained using the divergent kernel.  The only notable difference
is that the softened kernel gives a non-zero surface charge which
is replaced by a logarithmically divergent bound charge density at
the boundaries with no surface charge for the divergent kernel.
These are indistinguishable for all practical purposes.

Once the charge density is computed, the potential along the tube
is determined by
\begin{equation}\label{eq12}
U(z) = \int_0^L V(z-z^\prime) \rho(z^\prime) dz^\prime \hspace{1 mm}
.
\end{equation}
After expressing all lengths in terms of the tube radius and
integrating by parts, the potential becomes
\begin{equation}\label{eq13}
U(x) = \sum_i \frac{e_i}{R}\int_0^{\lambda_1}\!\!\!g_i(x^\prime)
\frac{\partial V (x-x^\prime)}{\partial x^\prime} dx^\prime
\hspace{1 mm} .
\end{equation}
Differentiating this expression and writing all constants in terms
of intensive variables allows us to relate the potential to the
strain through
\begin{equation}\label{eq14}
\frac{\partial U(x)}{\partial x} = \frac{e_2}{\lambda_2
\sin(3\theta)} (\beta_s - g_s(x))\hspace{1 mm} .
\end{equation}
Integrating this expression give the potential as an integral over
the strain field,
\begin{equation}\label{eq15}
U(x) = \frac{e_2}{\lambda_2 \sin(3\theta)} \int_x^{\lambda_1/2}
\!\!\!dx^\prime (g_s(x^\prime)- \beta_s)\hspace{1 mm} ,
\end{equation}
where we have set the potential to zero at middle of the tube.

Once $g_t(z)$ is obtained through Eq.~(\ref{eq08}), the change in
the azimuthal angle, $\phi$, at a point $z$ is given by $R d\phi =
g_t (z) dz$. The total twist induced along the tube is therefore
given by
\begin{equation}\label{eq16}
\Delta \phi = \int_0^{\lambda_1}\!\!\! dx \, g_t(x) \hspace{1 mm},
\end{equation}
where we have transformed to the scaled length.  Following the usual
convention, the total twist is positive if the tube rotates in the
counterclockwise direction, and negative if it rotates in the
clockwise direction.  Notice from Eq.~(\ref{eq08}) that for a
general chiral tube there is a coupling between longitudinal and
torsional strain such that a uniaxial stress alone ($\beta_t = 0$)
is sufficient to induce twisting in the nanotube. This coupling
vanishes in the high symmetry zigzag and armchair nanotubes, as may
be seen by writing Eq.~(\ref{eq08}) as $\lambda_2 \sin(3\theta) (g_t
(x)-\beta_t) = \lambda_3 \cos(3\theta) (\beta_s - g_s(x))$. For
armchair nanotubes $\sin(3\theta)=0$, which forces $g_s(x)=\beta_s$
but sets no restrictions on $g_t(x)$. Likewise, for zigzag nanotubes
$\cos(3\theta)=0$, which forces $g_t(x)=\beta_t$ but sets no
restrictions on $g_s(x)$.  For a general chiral tube the origin of
this ``stretch-twist'' coupling lies in the electrostatic term in
the free energy.  In regions where one flavor of strain has large
spatial gradients the bound charge density due to that flavor of
strain is also large.  The overall bound charge density may be
lowered in that region by gradients in the other flavor of strain
that produce bound charges of opposite sign. The equilibrium strain
fields are those that create a bound charge density profile that
gives the lowest electrostatic free energy consistent with
Eqs.~(\ref{eq08}) and (\ref{eq09}).

If we consider a system with $\beta_t = 0$, then Eqs.~(\ref{eq08})
and (\ref{eq16}) give
\begin{equation}\label{eqn01}
\Delta \phi = C \frac{\cos(3\theta)}{\sin(3\theta)},
\end{equation}
where $C=(\lambda_3/\lambda_2)\int(\beta_s-g_s(x))dx$.  We will see
below that for a system under tensile stress $\beta_s \geq g_s(x)$
for all $x$.  In this case $C \geq 0$, and the sign of the total
induced twist in the nanotube is entirely determined by the sign of
the chiral angle (mod $2\pi/3$).  We find that $\Delta \phi >0$ for
right-handed nanotubes and $\Delta \phi <0$ for left-handed
nanotubes.  Therefore, the sign of the induced twist may be used to
determine the handedness of a chiral piezoelectric nanotube. There
is no known order parameter for the handedness of a chiral nanotube
that can be built up out of the two primitive vectors of the $2$-d
hexagonal lattice and the translational vector of the nanotube.  As
such, an unambiguous assignment of nanotube handedness is difficult
with structural information alone.  A measurement of the sign of the
stretch-twist coupling allows one to easily make this assignment.

A non-piezoelectric chiral nanotube will also couple stretch and
twist to some degree due to the rearrangement of the atomic basis
upon stretching.  This effect is absent in our formalism because we
have set $C_{st} = 0$, when in reality this quantity is small but
non-zero for a chiral nanotube.  It is worth noting that the
electrostatically induced stretch-twist coupling described above is
in general several orders of magnitude larger than the coupling in
non-piezoelectric nanotubes.  The torsional strain induced in a
piezoelectric system can easily be of the same order of magnitude as
the uniaxial strain.

\section{\label{sec3} Equilibrium Solutions}

The equilibrium solutions for the strain fields $g_i(z)$ (Eqs.
\ref{eq08} and \ref{eq09}), the bound charge density $\rho(z) =
-\sum_i e_i \partial \eta_i (z)/\partial z$, and the electrostatic
potential $U(z)$ (Eq. \ref{eq15}) were determined using finite
element analysis to solve Eq. (\ref{eq09}).  Convergence of the
solution was monitored by increasing the number of finite elements
in increments of $100$ until the difference in successive solutions
was less than $1\%$ at the end of the tube where the convergence is
the slowest.

\subsection{\label{subsec3.1} Elastically Dominated Limit}

\begin{figure}
\includegraphics[width = 86mm]{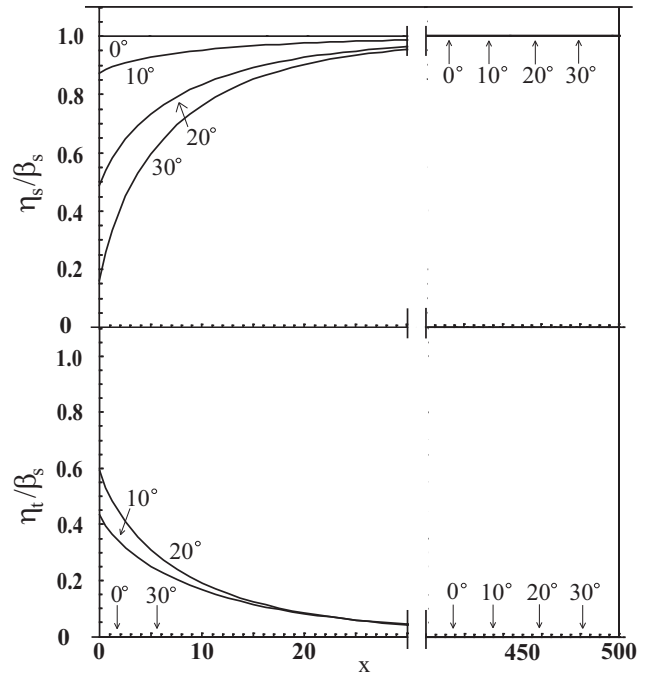}
\caption{\label{fig:estrains} Equilibrium solutions for a
piezoelectric tube in the elastically dominated limit using the soft
kernel. The upper panel shows the induced uniaxial strain for tubes
of various chirality, while the lower panel shows the corresponding
induced torsional strains.  The graph is broken along the $x$-axis
in order to show details near the tube boundary.  The graphs were
generated using $\beta_s = 1$,$\beta_t = 0$, $\lambda_1 = 1000$,
$\lambda_2 = 10/\pi$, $\lambda_3 = 2 \lambda_2$,and $e_2 =
\sqrt{5}/\pi$ (charge/length).}
\end{figure}

The elastically dominated limit is defined by the condition that
$2\pi\lambda_2$,$2\pi\lambda_3 \ll \lambda_1$.  We investigate the
case of a tube under constant uniaxial stress in the absence of
torsional stress. The equilibrium strain fields for several chiral
angles are presented in Fig.~\ref{fig:estrains}.  The strain fields
rapidly relax to their elastic limits, and over much of the tube
$\eta_s (x) \approx \beta_s$ and $\eta_t (x) \approx 0$. Near the
boundaries of the tube the electrostatic interactions effectively
increase $C_{11}$, suppressing the longitudinal strain. Importantly,
notice that for chiral tubes $\eta_t(x) \not= 0$ for all $x$ even
though $\beta_t = 0$.  Because the longitudinal and torsional
strains couple through spatial derivatives the chiral tubes develop
some nonzero twist near the tube ends.  The total twist induced in
the tube reaches a maximum at $\theta = 15^\circ$ and goes to zero
as $\theta \rightarrow 0^\circ$ or $\theta \rightarrow 30^\circ$.

\begin{figure}
\includegraphics[width = 86mm]{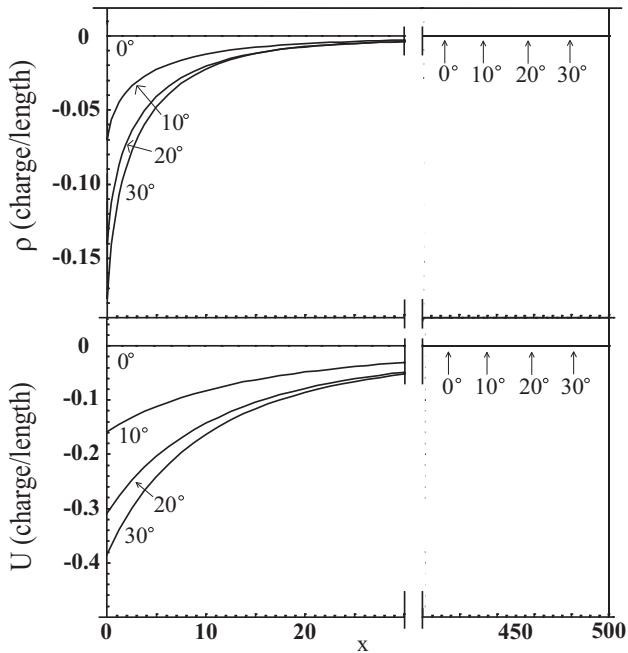}
\caption{\label{fig:epot} Equilibrium solutions for a piezoelectric
tube in the elastically dominated limit using the soft kernel. The
upper panel shows the induced bound charge density for various
chiral angles and the lower panel shows corresponding electrostatic
potential along the tube.  The graph is broken along the $x$-axis in
order to show details near the tube boundary.  The graphs were
generated using $\beta_s = 1$,$\beta_t = 0$, $\lambda_1 = 1000$,
$\lambda_2 = 10/\pi$, $\lambda_3 = 2 \lambda_2$,and $e_2 =
\sqrt{5}/\pi$ (charge/length).}
\end{figure}

The bound charge density and the electrostatic potential are plotted
in Fig.~\ref{fig:epot}.  The charge density is localized at the ends
of the tube and decays rapidly on the interior.  Not shown are delta
functions in the charge density right at the tube boundary. These
are surface charges which appear because the polarization is nonzero
at the tube boundary with the soft kernel. The potential varies
rapidly near the tube ends due to the presence of the induced bound
charges.  On the interior of the tube the potential varies slowly
and resembles the electrostatic potential produced by two opposite
point charges at the ends of the tube.  Almost the entire potential
drop across the tube occurs near the tube boundaries.

Both the bound charge density and the potential are maximized by
letting $\theta = 30^\circ$ and decrease monotonically as $\theta
\rightarrow 0$.  We find that the total potential difference
across the tube is well fit by $\Delta U \sim \sin(3\theta)$.
Therefore, under longitudinal stress, the piezoelectric response
is insensitive to small perturbations around the zigzag
configuration.

A striking feature of Figs.~\ref{fig:estrains} and \ref{fig:epot} is
that the effects of the chirality only manifest themselves at the
ends of the tube.  The bulk of the tube responds to an applied
stress as an ordinary elastic medium. This is a somewhat surprising
result because it implies that boundaries are necessary for chiral
effects to appear and that the bulk of the tube has no ``knowledge''
of its own chirality. Although this is evident from our free energy
formalism an explanation of this effect based on the symmetry
properties of the tube is desirable and is the subject of current
investigations.

An approximate analytic solution to Eq.~(\ref{eq09}) may be
obtained with a perturbation expansion and gives
\begin{equation}\label{eq17}
g_s (x) = \beta_s - (\gamma \beta_s - \tau)
\left[\frac{1}{(x+\alpha)^2} + \frac{1}{(\lambda_1 - x +
\alpha)^2}\right] \hspace{1 mm} .
\end{equation}
This expression always gives the correct asymptotic decay to the
elastic limit far from the tube ends.  If the system is very far
into the elastic limit, so that $(2\pi \lambda_2/\alpha^2) \ll 1$,
then this expression holds over the entire length of the tube.
Using Eq.~(\ref{eq15}) to compute the potential difference,
$\Delta U = U(\lambda_1) - U(0)$, across a tube under uniaxial
stress only gives
\begin{equation}\label{eq18}
\Delta U = C_U - \frac{4 \pi e_2 \sin(3\theta) \beta_s}{\lambda_1
+ \alpha} \hspace{1 mm}.
\end{equation}
The constant $C_U$ can be determined numerically.  If the system
is extremely elastic ($(2\pi \lambda_2/\alpha^2) \ll 1$), then
$C_U= 4 \pi e_2 \sin(3\theta) \beta_s/\alpha$.

We may use Eq.~(\ref{eq08}) to obtain $g_t(x)$ from $g_s(x)$, and
then compute the total twist along the tube using
Eq.~(\ref{eq16}).  For a tube under uniaxial stress only, this
gives
\begin{equation}\label{eq19}
\Delta \phi = C_{\phi} - \frac{4 \pi \lambda_3 \sin(3\theta)
\cos(3\theta) \beta_s}{\lambda_1 + \alpha} \hspace{1 mm}.
\end{equation}
The constant $C_{\phi}$ can be determined numerically.  If the
system is extremely elastic ($(2\pi \lambda_2/\alpha^2) \ll 1$),
then $C_{\phi}= 4 \pi \lambda_3 \sin(3\theta) \cos(3\theta)
\beta_s/\alpha$.

Equations (\ref{eq18}) and (\ref{eq19}) show that, for systems in
the elastic limit, the induced potential and the total twist
essentially do not scale with tube length.  This is a consequence
of the fact that the bound charge density is localized near the
tube ends with a decay length set by material parameters,
independent of length.

\subsection{\label{subsec3.2} Electrostatically Dominated Limit}

\begin{figure}
\includegraphics[width = 86mm]{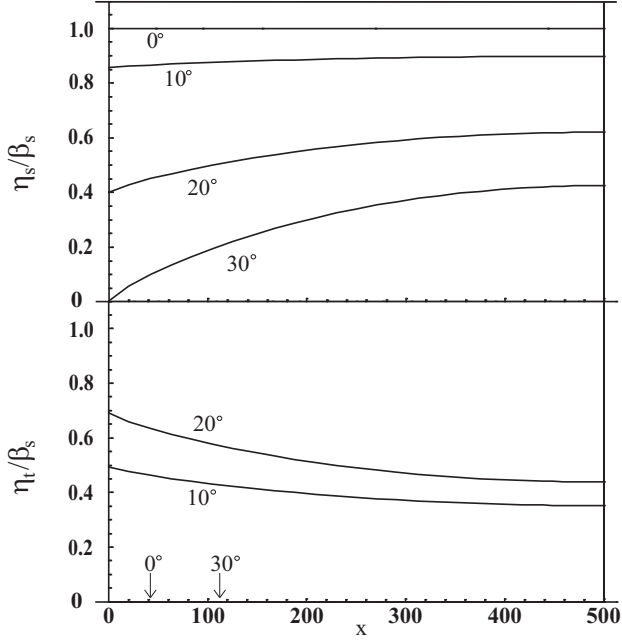}
\caption{\label{fig:cstrains} Equilibrium solutions for a
piezoelectric tube in the electrostatically dominated limit using
the soft kernel. The upper panel shows the induced uniaxial strain
for tubes of various chirality, while the lower panel shows the
corresponding induced torsional strains.  The graphs were generated
using $\beta = 1$, $\lambda_1 = 1000$, $\lambda_2 = 10000/\pi$,
$\lambda_3 = 2 \lambda_2$,and $e_2 = \sqrt{5000}/\pi$
(charge/length).}
\end{figure}

The electrostatically dominated limit is defined by the condition
$2\pi\lambda_2$, $2\pi\lambda_3 \gg \lambda_1$.  Again, we
investigate a tube under uniaxial stress only ($\beta_t = 0$).  In
Fig.~\ref{fig:cstrains} we plot the equilibrium strain fields for
several chiral angles.   Note that in this case the tube is shorter
than the strain decay length, and the strain fails to relax to its
elastic limit.  The tube develops significant torsional strain along
its entire length unless it is in one of the high symmetry
configurations.

\begin{figure}
\includegraphics[width = 86mm]{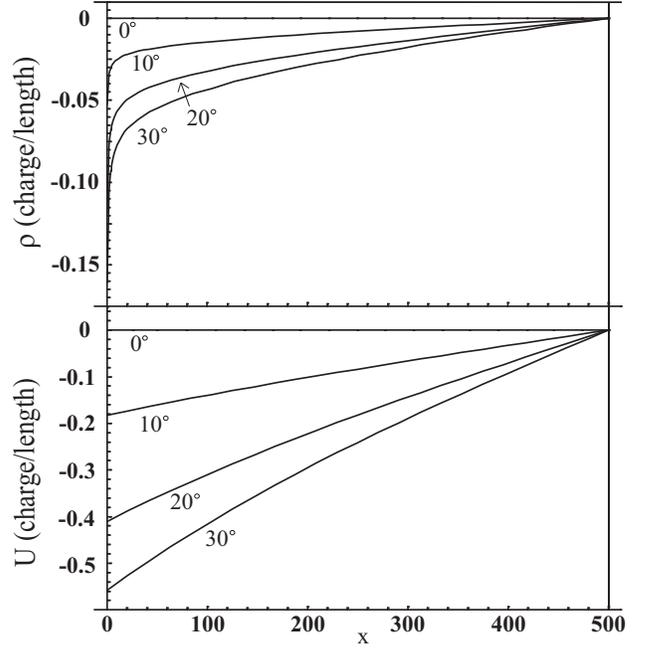}
\caption{\label{fig:cpot} Equilibrium solutions for a piezoelectric
tube in the electrostatically dominated limit using the soft kernel.
The upper panel shows the induced bound charge density for tubes of
various chirality, while the lower panel shows the corresponding
electrostatic potential along the tube. The graphs were generated
using $\beta = 1$, $\lambda_1 = 1000$, $\lambda_2 = 10000/\pi$,
$\lambda_3 = 2 \lambda_2$,and $e_2 = \sqrt{5000}/\pi$
(charge/length).}
\end{figure}

The bound charge density and the electrostatic potential in the
electrostatic limit are plotted in Fig.~\ref{fig:cpot} for several
chiral angles.  The bound charge density is nearly linear throughout
the interior of the tube and vanishes at the middle of the tube by
symmetry.  Near the boundaries the charge density begins to develop
a logarithmic divergence which is cut off because of the soft
kernel.  Not shown are delta functions in the charge density right
at the tube boundary.  Again, these are surface charge densities
that arise due to our use of the soft kernel.  The potential is also
nearly linear along the entire length of the tube and vanishes at
the center due to symmetry.  Near the boundaries there are small,
non-divergent corrections to the linear behavior.  A linear
electrostatic potential profile is what we would naively expect from
the conventional $3$-d case, but note the remarkably different
conditions which give rise to the linear potential in this case.  In
the $3$-d case the linear potential is produced by true surface
charges interacting via a long-range electrostatic interaction.  In
this $1$-d case, the linear potential is produced by a nearly linear
bound charge density interacting via a short-range electrostatic
interaction.

Inspection of the strains in Fig.~\ref{fig:cstrains} show they may
be well fit by quadratic curves.  An approximate analytic solution
may be obtained in the form $g_s(x) \approx c_o + c_2
(x-\lambda_1/2)^2$, where
\begin{eqnarray} c_0 & = & \left(\frac{1}{D}\right)
\left\{\frac{\beta_s \lambda_1^6}{180} + \left(5 \beta_s \gamma -
\tau\right) \left[\frac{\lambda_1^5}{60 \alpha} + O(\lambda_1^4
\log \lambda_1)\right] \right. \nonumber \\ & & \left. + \gamma
\tau \left[\frac{4\lambda_1^3}{3 \alpha}
\log\left(\frac{\lambda_1}{\alpha}\right) - \frac{28
\lambda_1^3}{9} + O(\lambda_1^2 \log \lambda_1)\right]\right\},
\nonumber \\
\end{eqnarray}
\begin{eqnarray}
c_2 = \left(\frac{\tau - \beta_s
\gamma}{D}\right)\left[\frac{\lambda_1^3}{3 \alpha} +
O(\lambda_1^2 \log \lambda_1)\right] \hspace{1 mm} ,
\end{eqnarray}
and
\begin{eqnarray}
D & = & \frac{\lambda_1^6}{180} + \gamma
\left[\frac{\lambda_1^5}{15 \alpha} + O(\lambda_1^4)\right]
\nonumber \\ & &  + \gamma^2 \left[\frac{4\lambda_1^3}{3 \alpha}
\log\left(\frac{\lambda_1}{\alpha}\right) - \frac{28
\lambda_1^3}{9} + O(\lambda_1^2 \log \lambda_1)\right] \hspace{1 mm} . \nonumber \\
\end{eqnarray}
This expansion is accurate to within a few percent provided
$\gamma$,$\tau \gg \lambda_1$, which is generally true if
$\lambda_2$, $\lambda_3 \gg 1$.  The exception occurs if $\theta
\rightarrow \pi/6$, in which case $\tau \rightarrow 0$ for any
$\lambda_2$, $\lambda_3$.  Even in this case the expansion differs
from the numerical results only near the tube ends, and still
gives acceptable agreement with the numerical results for the
potential and total twist.  For the potential across the tube we
find
\begin{equation}
\Delta U =\left(\beta_s - c_0\right) \lambda_1 - \frac{c_2}{12}
\lambda_1^3 \hspace{1 mm} .
\end{equation}
Because $(\beta_s - c_0) \gg c_2$ the potential scales nearly
linearly with tube length, an effect anticipated from our
numerical results. The total twist along the nanotube will also
have a nearly linear scaling with length, provided the system
remains in the electrostatic regime.

\section{\label{sec4} Boron Nitride Nanotubes}

We now apply our theory to calculate the piezoelectric response of a
BNNT under uniform longitudinal tensile stress. The $2$-D elastic
constants of a BN sheet are given by\cite{hgbr98,dnim92} $C_{11}
\approx .30$~TPa~nm and $C_{66} \approx .10$~TPa~nm, and the
piezoelectric constant is given by\cite{sm03} $e_2 = .12$ e/Bohr.
Experimentally accessible forces are on the nano-Newton scale;  we
arbitrarily assume a force of $1$ nN. Typical tube dimensions are
$R=1$ nm and $L = 1$ $\mu{\rm m}$.

Using these values, we find that $\lambda_1 = 1000$, $\lambda_2 =
4.4 \times 10^{-3}$, $\lambda_3 = 1.3 \times 10^{-2}$, $\beta_s =
5.3 \times 10^{-4}$ and $\beta_t = 0$.  These numbers show that
BNNTs are an extremely elastically dominated system, and our
perturbation formulas given in Sec.~(\ref{subsec3.1}) apply.
Equation~(\ref{eq17}) agrees with the numerical results for the
strain to better than $0.1\%$, and Eqs.~(\ref{eq18}) and
(\ref{eq19}), with $C_U= 4 \pi e_2 \sin(3\theta) \beta_s/\alpha$
and $C_{\phi}= 4 \pi \lambda_3 \sin(3\theta) \cos(3\theta)
\beta_s/\alpha$, give the electrostatic potential and total twist
with even better agreement.

For all chiral angles the strain fields decay to within $0.5\%$ of
their elastic limits at a depth of $2R$, and are always within
$3\%$ of the elastic limit.  The interior of the tube develops no
significant twist and essentially responds as an achiral nanotube.
The total twist, in radians, induced in the system as a function
of chiral angle is $\Delta \phi (\theta) = 8.7 \times 10^{-5}
\sin(3\theta) \cos(3\theta)$.  The system develops significant
surface charges and a bound charge density that decays rapidly on
the interior of the tube, and the entire potential drop occurs
within a few radii of the tube ends. Using the values from above
and converting to SI units gives the potential across the tube as
a function of the chiral angle as $U(\theta) = 23.0
\sin(3\theta)$~mV.

Based on these calculations, BNNTs should develop a rather weak
piezoelectric response with almost no induced twist regardless of
chirality.  Physically, this is explained by the exceptionally large
elastic constants of the BN sheet, which are inherited by the
nanotube and make it a stiff system. BNNTs are so elastically
dominated that the predicted strain decay lengths, between $1$ and
$2$ tube radii, place this system near the limits of what can be
described by our continuum theory.

\section{\label{sec5} Conclusion}

In this paper we have developed and solved a continuum theory to
describe the piezoelectric effect in $1$-D tubular systems.  These
systems exhibit different physical properties than their bulk
counterparts.  Unconventional physical effects include the
development a bound charge density near the tube ends and a
non-linear electrostatic potential profile along the tube.
Furthermore, the piezoelectric response of these systems couples
longitudinal strain and torsional strain.  This coupling represents
a new degree of freedom which may be important in nanotube
characterization and nanoscale device applications.

It would be interesting to measure the non-linear potential
variation along a nanotube to verify the unconventional
piezoelectric effects described here. BNNTs are the model physical
system in which to realize these effects.  However, BNNTs tend to be
stiff structures for which the electrostatic contributions to the
elastic properties are expected to be small.

This work was supported by the Department of Energy under grant
DE-FG-84ER45118.

\bibliography{paperedit2bib}

\end{document}